\documentclass[preprint,12pt]{elsarticle}

 \usepackage{graphicx, subfigure}
\usepackage{epsfig}
\usepackage{epstopdf}
\usepackage{hyperref}
\usepackage{caption} 
\usepackage{float} 
\usepackage{amssymb}
\usepackage{amsmath}
\usepackage{array}
\usepackage{multirow}

\biboptions{sort}
\biboptions{compress}

\journal{Nuclear Instruments and Methods B}

\begin{document}

\begin{frontmatter}



\title {Characterization of neutron transmutation doped (NTD) Ge for low temperature sensor development}


\author[label1,label2]{S.~Mathimalar}
\author[label1,label2]{V.~Singh}
\author[label1,label2]{N.~Dokania}
\author[label3]{V.~Nanal\corref{cor1}}
\ead{nanal@tifr.res.in}
\cortext[cor1] {Tel.: +91-22-22782333; fax: +91-22-22782133.}
\author[label3]{R.G.~Pillay}
\author[label3]{S.~Pal}
\author[label4]{S.~Ramakrishnan}
\author[label5]{A.~Shrivastava}
\author[label6]{Priya~Maheshwari}
\author[label6]{P.K.~Pujari}
\author[label7]{S.~Ojha}
\author[label7]{D.~Kanjilal}
\author[label8]{K.C.~Jagadeesan}
\author[label8]{S.V.~Thakare}
\address[label1]{India based Neutrino Observatory, Tata Institute of Fundamental Research, Mumbai 400 005, India.}
\address[label2]{Homi Bhabha National Institute, Anushaktinagar, Mumbai 400 094, India.}
\address[label3]{Department of Nuclear and Atomic Physics, Tata Institute of Fundamental Research, Mumbai 400 005, India.}
\address[label4]{Department of Condensed Matter Physics and Material Science, Tata Institute of Fundamental Research, Mumbai 400 005, India.}
\address[label5]{Nuclear Physics Divison, Bhabha Atomic Research Centre, Mumbai 400 085, India.}
\address[label6]{Radiochemistry Division, Bhabha Atomic Research Centre, Mumbai 400 085, India}
\address[label7]{Inter University Accelerator Centre, New Delhi 110 067, India.}
\address[label8]{Isotope Production and Applications Division, Bhabha Atomic Research Centre, Mumbai 400 085, India.}

\begin{abstract}
Development of NTD Ge sensors has been initiated for low temperature (mK) thermometry in The India-based Tin detector (TIN.TIN).  NTD Ge sensors are prepared by thermal neutron irradiation of device grade Ge samples at Dhruva reactor, BARC, Mumbai.  Detailed measurements have been carried out in irradiated samples for estimating the carrier concentration and fast neutron induced defects. The Positron Annihilation Lifetime Spectroscopy (PALS) measurements indicated monovacancy type defects for all irradiated samples, while Channeling studies employing RBS with 2 MeV alpha particles, revealed  no significant defects in the samples exposed to fast neutron fluence of $\sim~4\times10^{16}/cm^2$.   Both PALS and Channeling studies have shown that vacuum annealing at 600~$^\circ$C for $\sim2$ hours is sufficient to recover the damage in the irradiated samples, thereby making them  suitable for the sensor development.  

\end{abstract}
\end{frontmatter}
\section{Introduction}
\label{}
Neutron Transmutation Doped (NTD) Ge sensors are commonly used as low temperature sensors in the rare decay events like neutrinoless double beta decay~\cite{Arnaboldi, pasca} and dark matter search~\cite{Martineau}. Thermal neutron capture by Ge atom leads to the production of dopants~\cite{Haller} like Ga, As and Se. This doping method is preferred over the melt growth method due to its homogeneity~\cite{Itoh}.  
The temperature dependence of the resistivity of a NTD Ge sensor is given by~\cite{pasca} 
\vskip -0.6cm
\begin{equation}
\rho = \rho_0 ~exp\left\lbrack \frac {T_0}{T}
\right\rbrack^{\alpha}
\end{equation}
 where the parameters $\rho_0$  depends on intrinsic properties of Ge,   T$_0$ depends on the doping level and the constant $\alpha$ is close to 0.5~\cite{Woodcraft}. It is important to achieve a high sensitivity (dR/dT) keeping resistance R of the sensor within the measurable range at mK temperature. The desired neutron fluence will depend on the actual isotopic composition of the Ge wafer. 
Since the reactor neutrons are not monoenergetic, samples are exposed to neutrons having a broad range of energies ($<$~0.625~eV to 10~MeV) during the irradiation. In particular, high energy neutrons can cause  damage to the lattice structure of the crystal. Even for thermal neutrons, the average recoil energy ($\sim$~182~eV) of Ge(n,$\gamma$) reaction is sufficiently high to drive the recoiling Ge atom into the interstitial position~\cite{Schweinler}. The dopant atom in interstitial positions cannot contribute to the total number of carrier concentration~\cite{Hasiguti}. Thus, defects created during the irradiation can significantly affect the performance of the sensor at low temperature. It should be mentioned that defect formation may depend on irradiation conditions, since  both the flux and energy of fast neutrons are responsible factors.
The performance of NTD Ge sensors over a wide range of carrier concentration 3~x~10$^{14}/cm^3$ to 1.8~x~10$^{17}/cm^3$ was studied in detail by Itoh~\textit{et al.}~\cite{ItohJLTP}.
Generally high temperature annealing is employed to cure the crystal defects and annealing conditions need to be optimized depending on the defect type and the density. 
 Kuriyama \textit{et al.}~\cite{Kuriyama} have reported that 30~min annealing in the Ar gas at 600$^\circ$~C  for a hole concentration of 3.5~x~10$^{14}/cm^{3}$ was sufficient to recover damages, while Pasca~\textit{et al.}~\cite{pasca} have employed annealing at 400$^\circ$~C for 6~hours for  carrier concentration in the range of 5~--~6.8~x~10$^{16}/cm^3$. 
Palaio ~\textit{et al.}~\cite{palaio} have also discussed effects of thermal annealing  in the Ar gas and have reported that annealing at higher temperatures (500$^\circ$~C) resulted in  contamination of samples due to rapidly diffusing impurities.  They further observed that the NTD Ge samples left at ambient temperature for eight months after irradiation showed improvements similar to thermal annealing. This is desirable for rare decay studies as annealing can introduce trace impurities in the sensors.

Development of NTD Ge sensors for mK thermometry in cryogenic bolometer detector TIN.TIN (The India-based TIN detector) to search for neutrinoless double beta decay in $^{124}$Sn  has been initiated~\cite{nanal, mathi-wolte}.  NTD Ge sensors are prepared, by neutron irradiation of device grade Ge samples at Dhruva reactor, Bhabha Atomic Research Centre (BARC), Mumbai, for the first time. Initial results for a NTD Ge sensor performance at 100~mK  have been reported in Ref.~\cite{mathi-wolte}. A systematic study of the fast neutron induced damage in the NTD Ge samples  for neutron fluence in the range  of 2~x~$10^{18}/cm^2$ to 1.4~x~$10^{19}/cm^2$ which corresponds to the carrier concentration of (0.5 - 3.3)~x~10$^{17}/cm^3$ is presented in this paper.
The neutron induced damage is investigated using Positron Annihilation Lifetime Spectroscopy (PALS) and Channeling in irradiated samples before and after annealing. 
In this paper Section II describes the experimental details, data analysis and results are presented in Section III. Conclusions are given in Section IV.

\section{Experimental details}
For sensor development two sets of semiconductor grade Ge samples from University wafers with  $<111>$ (0.4~mm thick, $\rho~\sim 30 ~\Omega~cm$, 5N purity, n-type)   and $<100>$ (1 mm thick, $\rho~\sim~35~\Omega~cm$,  5N purity, n-type) cleavage plane were used. The Ge samples were single side mirror polished.
The cleavage plane $<100>$ is preferred over the $<111>$ in terms of clean cutting of smaller sensors from the irradiated sample. Actual isotopic abundances of both $<100>$ and $<111>$ samples were obtained with the Secondary Ion Mass Spectrometry (SIMS)~\cite{mathi-nim}. It was observed that $^{70}$Ge and  $^{74}$Ge, which produce $^{71}$Ga and $^{75}$As, respectively, are fractionally more in  $<100>$ compared to that in $<111>$ and hence, the net carrier concentration  for a given neutron flunece is expected to be similar. The production rate of dopants are estimated from the measured isotopic abundance and n-capture cross section taken from Ref~\cite{sigma}. These estimated production rates per unit neutron fluence and the expected net carrier concentration values are tabulated in Table~\ref {table5}.
The Scanning Electron Microscope (SEM) and Energy Dispersive X-ray analysis (EDAX) studies of the Ge wafer prior to irradiation showed no surface impurities other than traces of oxygen. 
The neutron irradiation was carried out at Dhruva reactor, BARC, Mumbai (India) for samples of different sizes (10~x~30~mm$^2$ or 10~x~10~mm$^2$) in different batches. 
Required neutron fluence was estimated based on measured isotopic abundances but actual neutron fluence varied depending on the reactor power. Typically, samples were irradiated for 4~-~6 days and the estimated thermal as well as fast neutron fluence are given in Table~\ref{table1}. A minimum cooldown period of 45 days was required before the irradiated samples could be handled for measurements. Although the samples were cleaned with electronic grade alcohol prior to irradiation (as described in Ref.~\cite{mathi-nim}), the irradiated samples showed lack of lustre. Hence, the irradiated samples were etched with HF (40$\%$) to clean the surface. It should be mentioned that all measurements have been carried out 4 to 8 months after the irradiation.

\begin{table}[h!]%
\caption{ { Estimated production rate of dopants ($atoms/cm^3$) per neutron, obtained by multiplying measured isotopic abundance of Ge~\cite{mathi-nim} with n-capture cross section (taken from Ref.~\cite{sigma}). Net carrier concentration (p-type) = $^{71}$Ga - ($^{75}$As + 2 $\times$ $^{77}$Se) is given in the last column.}}
    \centering  
    \label{table5}
 \begin{tabular}{ccccc}
\hline
Sample & $^{71}$Ga &  $^{75}$As &  $^{77}$Se & Net carrier \\
&&&&concentration \\
\hline
$<$111$>$ & 3.23~$\times$~$10^{-2}$ &  7.78~$\times$~$10^{-3}$ &  4.76~$\times$~$10^{-4}$ &  2.35~$\times$~$10^{-2}$\\
$<$100$>$ & 3.32~$\times$~$10^{-2}$ &  7.89~$\times$~$10^{-3}$ &  5.08~$\times$~$10^{-4}$ &  2.43~$\times$~$10^{-2}$\\
\hline
\\
\end{tabular}
\end{table}

  \vskip -0.2cm
\begin{table}[h!]%
\caption{Details of estimated thermal and fast neutron fluence from average reactor power during irradiation.}
    \centering  
    \label{table1}
  \small\addtolength{\tabcolsep}{-1.7pt}
 \begin{tabular}{ccccc}
\hline
\multirow {4} {*}{Sample} &  Cleavage   & \multicolumn{3}{c}{Neutron fluence~($n/cm^{2}$)}\\
 &plane &  Thermal& Epithermal &Fast\\ 
 & &($<$~0.625~eV)&  (0.625~eV to  &(0.821 to  \\
 &&& 0.821~MeV)& 10MeV)\\
\hline
B  & $<$111$>$ &1.40 x 10$^{19}$ & 2.56 x 10$^{18}$ & 1.72  x 10$^{17}$\\ 
C  & $<$111$>$ & 9.13 x 10$^{18}$ & 1.72 x 10$^{18}$ & 1.12 x 10$^{17}$ \\ 
E & $<$100$>$ & 2.11 x 10$^{18}$ & 3.52 x 10$^{17}$ & 2.46 x 10$^{16}$\\
F  & $<$100$>$ & 3.52 x 10$^{18}$ & 5.87 x 10$^{17}$ & 4.11 x 10$^{16}$\\
\hline
\multicolumn{4}{l} { expected errors in reactor power are within 0.2 $\%$.}
\\
\end{tabular}
\end{table}
\vskip -0.1 cm
In the present case,  the thermal neutron flux  covered a higher range than the reported values in Ref.~\cite{Kuriyama}. Hence, the  annealing was done for longer duration, namely, at 600$^o$~C for 2~hours.  Samples were annealed in the vacuum sealed quartz tube to avoid contamination. 

  Carrier concentrations in annealed samples were estimated from the Hall effect measurement. Ohmic contacts were made on NTD Ge sample by depositing Au(88\%)--Ge(12\%) (pattern of six pads) of about 50~nm thickness and rapid thermally annealing at 400$^o$C for 2~min. The Au-Ge contact pad were wedge bonded with Aluminum wires ($\phi$ = 25~$\mu$m) to the mounting chip for making the electrical connections. The measurement was carried out using Van der Pauw method and by varying the magnetic field from --1~T to +1~T at 77~K and 300~K.  Defect studies were carried out  using PALS and Channeling in both unannealed and annealed samples. 
The PALS measurements were done with $^{22}$Na  positron source at Radio Chemistry Division, BARC. The Channeling measurements were performed with 2~MeV alpha beam from Pelletron Accelerator RBS-AMS Systems  facility (PARAS) at Inter University Accelerator Centre (IUAC), NewDelhi~\cite{paras}. Details of measurement techniques and sample preparation are described in Ref.~\cite{mathi-wolte}.

\section{Data Analysis and Results}
\subsection {Hall Effect Measurement}
\begin{figure}[h!]

          \centering
          \includegraphics[scale=0.4]{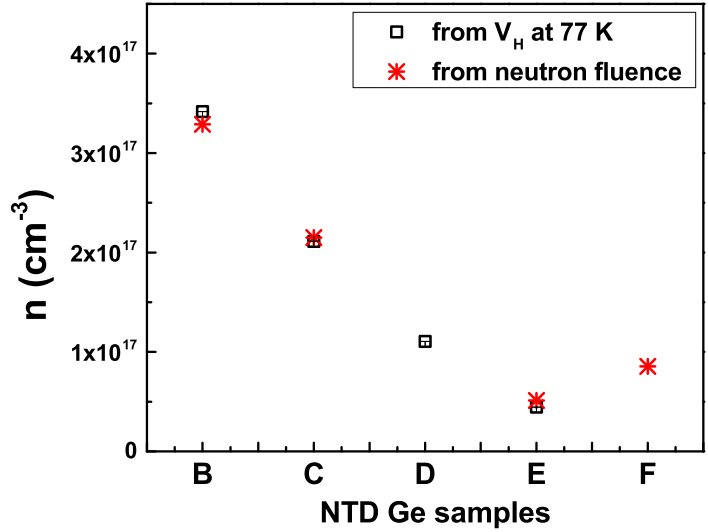}
        \hspace{0.1cm}
               
        \caption{ (color online) The measured carrier concentration (n) of NTD Ge from Hall effect at 77~K along with expected carrier concentration based on the given neutron fluence.}
        \label{fig1}
      \end{figure}

From the measured Hall voltage V$_H$, the carrier concentration~(n) is estimated using the following expression:
\begin{equation}
n =   \left\lbrack 
\frac {B . I}{e.V_H.t}
\right\rbrack
\end{equation}
where B is the applied magnetic field, I is supplied current  and t is the thickness of the sample. It should be mentioned that the ratio $V_H/B$ is averaged over magnetic field range --1~T to +1~T, in order to eliminate the contribution from the magneto-resistance.   The measured carrier concentration at 77~K and 300~K are tabulated in Table~\ref{table2}. The carrier concentrations expected from the neutron fluence, which in turn are derived from the reactor power, are also listed in Table~\ref{table2}.  As explained by  Morin~\cite{Morin} carrier concentration estimated at 77~K, where the Hall factor is close to unity, should be used for comparison with values calculated from the neutron fluence and is plotted in Figure~\ref{fig1}.
It can be seen from the Figure~\ref{fig1} that the carrier concentration obtained from the Hall effect measurement and the neutron fluence shows good correlation.  Data for samples D~\cite{mathi-wolte} and F is also shown to indicate the complete range. 
\begin{table}[h!]
\centering
\caption{ The carrier concentrations of NTD Ge from Hall effect measurement at 77~K and 300~K along with expected carrier concentrations obtained from the given neutron fluence. Net carrier concentration (p-type) = $^{71}$Ga - ($^{75}$As + 2 $\times$ $^{77}$Se) is given in last column.}
\label{table2}
\begin{tabular}{lrccccc}
\hline 
\multirow {3} {*}{Sample}  &\multicolumn{6}{c}{Carrier concentrations ($\times 10^{16}cm^{-3}$)}\\
& \multicolumn{2}{c} {Hall Measurement} & \multicolumn{4}{c} {From neutron fluence} \\
& 77K* & 300K* & $^{71}$Ga & $^{75}$As & $^{77}$Se & Net carrier \\
&&&&&&concentration\\
\hline
B &34.20 & 38.90  & 45.16 & 10.89 & 0.67 & 32.93  \\
C& 21.10 & 26.00  & 29.45 & 7.10 & 0.44 &21.47\\
E & 4.45  & 7.25 & 7.00 & 1.66 & 0.11 &5.12\\
\hline 
\multicolumn{7}{l} {* Errors are within 0.1 $\%$.}  \\
\end{tabular}
\end{table}


In the present data for  $<100>$ samples  (sample D reported in Ref.~\cite{mathi-wolte} and sample E) Hall factor at 300~K was found to be closer to $\sim$~1.8 (value reported in 
 Ref.~\cite{Morin}), while $<111>$ samples (B and C) show a much smaller value close to $\sim$~1.2. 
 It should be mentioned that samples D, E and F have carrier concentration in the range of $0.5-1.0$~x~10$^{17}/cm^3$, which is suitable for low temperature thermometry below 1K. It should be mentioned that sample B with n~$\ge$ 3.3~$\times$~10$^{17}/cm^3$ has very low and nearly flat dR/dT at 100~mK. Hence this sample is unsuitable for mK thermometry.

\subsection {Defect studies}
  
 Table~\ref{table3} gives the PALS results of irradiated samples before and after annealing. The measured lifetime of positron ($\tau_{\beta^+}$) for both $<100>$ and $<111>$ virgin samples is consistent with that of the bulk Ge crystal ($\tau_{\beta^+}$~=~228~ps). The measured positron lifetime $\tau_{\beta^+}$~$\sim$~293.6~ps in all the irradiated samples before annealing,  indicates that defects are of  \lq{monovacancy}\rq~ type.  The intensity of the $\tau_{\beta^+}$ lifetime were 100$\%$ for all the samples except for the sample D, where  two lifetime components corresponding to monovacancy (Intensity~$\sim$~{90.0 $\pm$ 2.5} $\%$) and vacancy cluster (Intensity~$\sim$~{9.4 $\pm$ 2.4}$\%$) were observed~\cite{mathi-wolte}. The observed single lifetime implies that  $\beta^+$ has undergone saturation trapping. Hence, the exact defect concentration can not be inferred and is estimated to be greater than 0.1~$\%$. The $\tau_{\beta^+}$ observed for all the annealed samples is very similar to that of the virgin crystal indicating complete recovery of defects.
\begin{table}[h!]
\centering
\caption{\label{tab:table3} Measured positron lifetime ($\tau_{\beta^+}$) in NTD Ge using PALS.}
\label{table3}
\begin{tabular}{cccl}
\hline 
\multicolumn{2}{c}{Samples }  & $\tau_{\beta^+}$  & $\tau_{\beta^+}$~\cite{PALS}  \\
&& (ps) &(ps)\\ \hline
{virgin $<111>$} &   &{ 227.8 $\pm$ 0.3}  &{228}(Bulk)  \\
 {virgin $<100>$} & & {232.0 $\pm$ 0.3}  & {228}(Bulk) \\
{B}&  {irr.} &  {293.6 $\pm$ 0.4}  &{293}(Monovacancy)\\
& {ann.}  & {225.6 $\pm$ 0.3}  &{228}(Bulk)\\
{E}& {irr.}  &  {293.5 $\pm$  0.4}  &{293}(Monovacancy)  \\
& {ann.}  &{228.0 $\pm$ 0.4}  &{228}(Bulk) \\
 F&{irr. }   &{294.0 $\pm$  0.6} & {293}(Monovacancy)\\
{D \cite{mathi-wolte}}& {irr. } &  {294.0 $\pm$ 0.3}    &{293}(Monovacancy)  \\
&&  {401 $\pm$ 30} & {401}(Vacancy clusters) \\
&{ann.}  & {233.0 $\pm$ 0.4} &{228}(Bulk) \\

\hline
\multicolumn{4}{l} {irr. - irradiated}\\
\multicolumn{4}{l} {ann. - annealed}\\
\end{tabular}
\end{table} 

\begin{figure}[h!]
          \centering
          \includegraphics[scale=0.37]{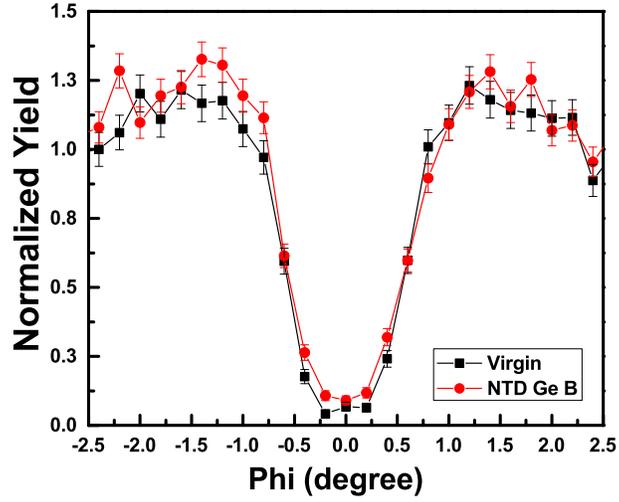}
          \caption{ (color online) The axial scan of irradiated NTD Ge sample B together with that for the virgin $<111>$ sample.}
          \label{fig2}
\end{figure}

\begin{figure}[h!]
          \centering
          \includegraphics[scale=0.4]{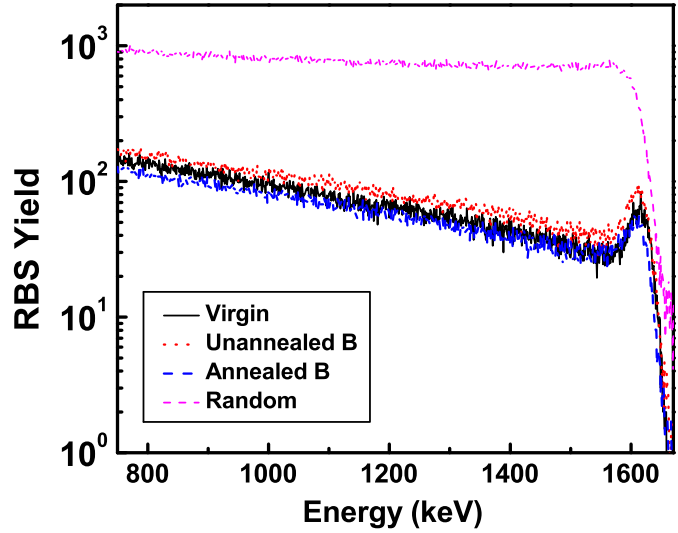}
  \caption{ (color online) The RBS spectra  along $<111>$ axis  of the unannealed and annealed Sample B  along with the virgin sample.  The spectrum in the random orientation is also shown for comparison.}
          \label{fig3}
\end{figure}

 It was found that the RBS spectra of samples E and F irradiated with lower neutron fluence ($N_{fast}  < 4 \times 10^{16}~  n/cm^2$) were very similar to that of the virgin sample. However for the higher neutron fluences, i.e. in sample D~\cite{mathi-wolte} and sample B, increase in the RBS yield of the channeled spectra was observed. The axial scan for n-irradiated $<111>$ sample is shown in Figure~\ref{fig2} together with that for the virgin sample. It can be seen that $\chi_{min}$ in NTD Ge B  is 0.09~$\pm$~0.02 is higher compared to that for the virgin sample ($\chi_{min}$ = 0.06~$\pm$~0.02). It is also evident that the measured half angle 0.55$^\circ$~$\pm$~0.05$^\circ$ is similar for both the samples. No measurable changes in the width  of the Channeling dip  were observed for any of the samples.
  Figure~\ref{fig3} shows the channeled spectra for the sample B after annealing and the virgin $<111>$ sample. It is evident that the channeling effect observed in the annealed sample B is somewhat better than the virgin sample, implying complete recovery of damages.
It should be mentioned that unannealed samples showed the presence of defects even after four months since irradiation. Thus, self-annealing at ambient temperature did not suffice and high temperature annealing was essential. 

The $\chi_{min}$~=~Yield (channeled)/Yield (random) is obtained by taking the yield for about 100~keV (corresponding to a depth of 316~nm) below the surface peak of the channeled spectrum.
 The defect density $N_D$ is estimated using the following equation~\cite{Satoh},
 \begin{equation}
N_D = N  \left\lbrack 
\frac {\chi_{min}^{irr}-\chi_{min}^{o}}{1-\chi_{min}^{o}}
\right\rbrack
\end{equation}
where $N$ is the number density of Ge atoms, $\chi_{min}^{irr}$ is the $\chi_{min}$ of the irradiated sample and $\chi_{min}^{o}$ is that of virgin sample. 

\begin{center}
\begin{table}[h!]%
\caption{Estimated $N_D$ in irradiated samples from the channeling data.}
\centering
\label{table4}
\begin{tabular}{cccc}
\hline
Samples & $\chi_{min}^{o}$ & $\chi_{min}^{irr}$ & $N_D$~x 10$^{20}$ \\ 
&(\%)&(\%)& $/cm^3$ \\\hline
B& 4.44 $\pm$ 0.08 & 5.92 $\pm$ 0.06& {(6.8 $\pm$ 0.4) } \\
 D&  4.19 $\pm$ 0.05 & 5.76 $\pm$ 0.06 &{(7.3 $\pm$ 0.4)} \\
\hline
\end{tabular}
\end{table}
\end{center}

  The estimated defect densities for samples B and D, where measurable change in $\chi_{min}$ was observed for unannealed samples with respect to virgin samples are tabulated in Table \ref{table4}. The defect concentration from the Channeling experiment is estimated to be 1.6~$\%$ and is consistent with PALS measurement~($>$~0.1~$\%$). It may be added that the estimated $N_D$ values for 50~keV and 100~keV energy window below the surface peak defined depth are similar within error, indicating no significant dechanneling effects in near surface region.

\vskip -0.9 cm
  
\section{Conclusions}
The development of NTD Ge sensors has been initiated for mK thermometry in TIN.TIN detector.  NTD Ge sensors are prepared, by thermal neutron irradiation of device grade Ge samples at Dhruva reactor, BARC, Mumbai, for the first time. The carrier concentration in irradiated, annealed samples was estimated from the Hall effect measurement at 77~K and is found to be in good agreement with that  estimated from the neutron fluence. Fast neutron induced defects have been studied using PALS and Channeling. The PALS measurements indicated monovacant type defects for all irradiated samples, whereas the  Channeling studies revealed no significant defects for 
fast neutron fluence upto $\sim~4~$x~10$^{16}/cm^2$. 
These fast neutron induced defects were observed in the irradiated samples even after about 4~-~8 months of irradiation. Thus unlike in Ref.~\cite{palaio},  for neutron fluences $>$~10$^{17}/cm^2$ studied in the present work leaving the samples at ambient temperature is not sufficient.
It is shown that vacuum annealing at 600~$^\circ$C for $\sim2$ hours is sufficient to recover the damage in the irradiated samples, as verified from both PALS and Channeling. The annealed samples with carrier concentration in the range $0.5-1$~x~10$^{17}/cm^3$ are found to be suitable for the sensor development. The sensor made from the annealed NTD Ge sample was tested in the temperature range 75 - 250 mK and the results are reported in the  Ref.~\cite{mathi-wolte}. 
\section{Acknowledgement}
We thank Ms. S. Mishra for help with sample preparation;   Mr. J. Mathew, Mr. A. Singh, Dr. S.S. Prabhu, Prof. M. Deshmukh for assistance  with  electrical measurements and  Ms. B.A. Chalke for SEM measurements. We are grateful to Dr. V.M.~Datar  and Dr. G.~Ravikumar for useful discussions.

\end{document}